\documentclass[acmsmall, screen, nonacm]{acmart}
\AtBeginDocument{%
  }
\setcopyright{acmlicensed}
\copyrightyear{2018}
\acmYear{2018}
\acmDOI{XXXXXXX.XXXXXXX}
\acmConference[Conference acronym 'XX]{Make sure to enter the correct
  conference title from your rights confirmation email}{June 03--05,
  2018}{Woodstock, NY}
\acmISBN{978-1-4503-XXXX-X/18/06}

\usepackage[inline]{enumitem}

\setlist{noitemsep}
\setlist[enumerate]{left=0pt .. \parindent, labelsep=5pt}

\begin{document}
\title[Survival of the Notable]{Survival of the Notable: Gender Asymmetry in Wikipedia Collective Deliberations}

\author{Khandaker Tasnim Huq}
\orcid{0000-0002-4622-8102}
\email{kthuq@umd.edu}
\affiliation{%
  \institution{University of Maryland}
  \city{College Park}
  \state{Maryland}
  \country{USA}
}

\author{Giovanni Luca Ciampaglia}
\orcid{0000-0001-5354-9257}
\email{gciampag@umd.edu}
\affiliation{%
  \institution{University of Maryland}
  \city{College Park}
  \state{Maryland}
  \country{USA}
}
\renewcommand{\shortauthors}{Huq and Ciampaglia}

\begin{abstract}

    Communities on the web rely on open conversation forums for a number of
    tasks, including governance, information sharing, and decision making.
    However these forms of collective deliberation can often result in biased
    outcomes. A prime example are Articles for Deletion (AfD) discussions on
    Wikipedia, which allow editors to gauge the notability of existing articles,
    and that, as prior work has suggested, may play a role in perpetuating the
    notorious gender gap of Wikipedia. Prior attempts to address this question
    have been hampered by access to narrow observation windows, reliance on
    limited subsets of both biographies and editorial outcomes, and by potential
    confounding factors. To address these limitations, here we adopt a competing
    risk survival framework to fully situate biographical AfD discussions within
    the full editorial cycle of Wikipedia content. We find that biographies of
    women are nominated for deletion faster than those of men, despite editors
    taking longer to reach a consensus for deletion of women, even after
    controlling for the size of the discussion. Furthermore, we find that AfDs
    about historical figures show a strong tendency to result into the
    redirecting or merging of the biography under discussion into other
    encyclopedic entries, and that there is a striking gender asymmetry:
    biographies of women are redirected or merged into biographies of men more
    often than the other way round. Our study provides a more complete picture
    of the role of AfD in the gender gap of Wikipedia, with implications for the
    governance of the open knowledge infrastructure of the web. 
    
\end{abstract}

\ccsdesc[500]{Human-centered computing~Empirical studies in collaborative and social computing}
\ccsdesc[300]{Social and professional topics~Gender}
\ccsdesc[100]{Mathematics of computing~Survival analysis}

\keywords{Wikipedia, Gender Gap, Collective Deliberation, Article for Deletion, Competing Risks Analysis}


\maketitle

\section{Introduction}

In 2018, the physicists Donna Strickland, Gérard Mourou, and Arthur Ashkin
received the Nobel Prize for ``groundbreaking inventions in the field of laser
physics''\footnote{Morou and Strickland shared half of the price for their work
on chirped pulse amplification; the other half went to Ashkin for separate
work.}. The Nobel prize in Physics recognizes a lifetime of accomplishments of a
scientific nature, but this award carried additional significance since it was
the first to go to a woman in 55 years. Nonetheless, in the immediate aftermath
of the announcement, internet users looking to know more about Donna Strickland
could not find a Wikipedia entry about her --- as one would normally expect when
looking up an accomplished scientist --- even though her two co-awardees had
been listed in the encyclopedia since 2005. Later, it was discovered that there
had indeed been an attempt to create an entry for her in Wikipedia. Some editors
had drafted her biography, but this draft had never been published, and thus had
never been indexed by search engines, since --- unlike her scientific work ---
she had not been considered to be `notable' enough to warrant the creation of a
separate, individual entry about her~\cite{Bazely18}. 

On February 11, 2019, the biography of nuclear scientist Clarice Phelps, who is
the first African American woman to help discover a new chemical element, was
deleted from Wikipedia~\cite{Jarvis19}. The debate over whether Phelps met the
internal notability standards of Wikipedia was contentious. Editors initially
deleted her entry, then reinstated it, only to delete it again later. In total,
her entry was deleted (and later reinstated) three times --- each deletion
sparking increasingly heated discussions. Last but not least, in 2017 a
Wikipedia editor raised concerns about the notability of Margaret D. Foster
(1895--1970), an American chemist known for being the first female chemist at
the United States Geological Survey and for her involvement on the Manhattan
Project~\cite{Harrison19}. Despite this record, her Wikipedia page was flagged
for potential deletion due to concerns about her notability.

These three scenarios --- which are drawn from STEM areas but are not
necessarily limited to those fields --- exemplify how barrier-breaking women get
undermined and undervalued and face challenges to be fully included in Wikipedia
--- the largest and most influential online encyclopedia. Though this issue
reflects broader challenges women encounter across the web, the central role of
Wikipedia in shaping public knowledge makes it a critical site where women
struggle for the recognition they deserve. 

As a free encyclopedia, Wikipedia aspires to encapsulate the entirety of human
knowledge. It stands as a significant source of encyclopedic information
encompassing notable individuals from diverse countries, historical periods, and
fields of knowledge on a global scale. Within its extensive body of knowledge,
about 1.9 million entries cover the biographies of notable individuals like
scientists, artists, politicians, etc., yet very few of these biographies are
about women (approximately 19\% in the English version, the largest of all
Wikipedia versions~\cite{konieczny2018gender}), highlighting a substantial
gender gap in encyclopedic coverage between men and women. This is concerning,
since the recognition of the merits of an individual, particularly in a societal
context, requires fair representation, but also because the presence of this
kind of gap in one of the pillars of the digital knowledge infrastructure poses
the risk of propagating preexisting social biases in other digital platforms,
and thus of further strengthening the underlying gender inequality in
society~\cite{conway2018disappeared}. For example, Wikipedia is often used to
train Machine Learning models for a variety of tasks, representing one of the
highest quality corpora openly available. As a result, several AI models are at
risk of being biased against women in various
applications~\cite{stanovsky2019evaluating, gor2021deconfounding,
kotek2023gender}.

\begin{figure}
  \centering
  \includegraphics{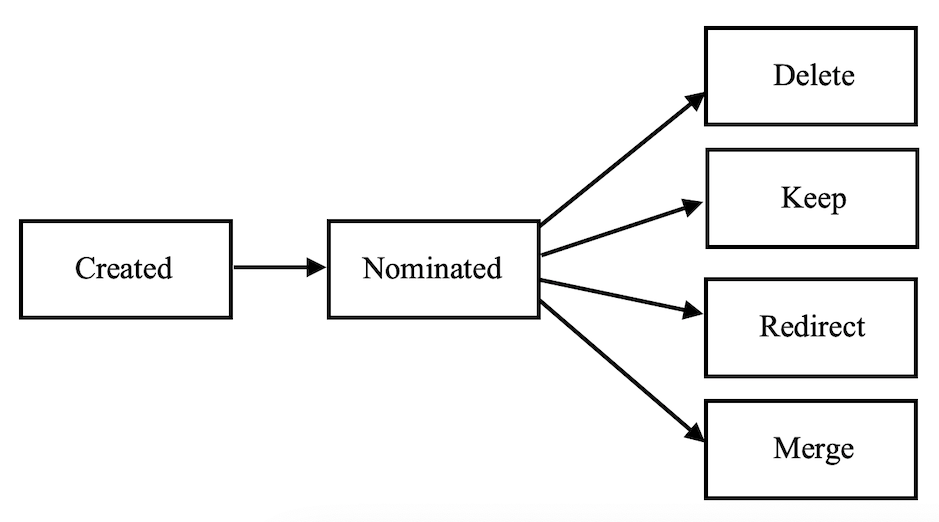}
  \caption{Multi-state model depicting competing risks of nomination and each outcome of the deliberation}
 \label{fig:diagram}
  \Description{Multi-state model depicting competing risks of nomination and each outcome of the deliberation}
\end{figure}

Prior work on the gender gap in Wikipedia has found that articles about women
are brought up for deletion more frequently than those about
men~\cite{tripodi2023ms,martini2023notable, lemieux2023too}, yet it is not clear
how and when this takes place. Like many other editorial actions, the deletion
of an article is subject to the regular collaborative norms of
Wikipedia~\cite{Reagle2007}, and thus articles are susceptible of being flagged
for deletion at any point during their editorial cycle. Our objective is thus to
investigate the following question: how quickly do Wikipedia biographies
interact with the Articles for Deletion (AfD) process? Specifically, we aim to
determine if there is any gender-based discrepancy in the various stages of the
process, from the initial nomination for deletion to the consensus outcome taken
by the discussants (if any). Our analysis covers the full history of the AfD
process, from January 15, 2001 to November 3, 2023. 

\paragraph{Definition of gender} For our analysis, we rely on human annotations provided by the Wikidata project
to identify the gender of biographical subjects. Even though the vital records
information from Wikidata provided us with rich gender information about the
subjects of biographies, in this study, we chose to restrict ourselves to a
gender binary (i.e., man and woman). This choice was dictated by the
limited frequency of other, non-binary gender labels in the dataset (0.09\%),
which would have severely limited our ability to draw reliable statistical
inferences about them. Thus, it is important to acknowledge that our analysis is
restricted in its scope, as it does not encompass all existing notions of
gender.

The rest of the article proceeds as follows. First, to define the scope of the
problem, we review how prior work has documented the gender gap on Wikipedia and
its community. We then provide a brief description of how the Wikipedia deletion
process works and discuss prior literature on gender and deletion discussions.
From this discussion we derive the hypotheses that guide our main research
question, and highlight how a host of contingent factors, like the historical
evolution of Wikipedia as a mass collaboration project, are relevant when
measuring the gender disparity in deletion discussions. Since Wikipedia is a
repository of both historical and contemporary knowledge, we also considered two
more factors: the additional considerations that Wikipedians typically take into
account when debating about the merits of living people to be featured on
Wikipedia, and the challenges in finding information about historical figures.
In the Discussion section, we go back to the motivating research question,
consider the main limitations of our analyses, and point to ways in which future
work can build on our approach.

\section{Background}

While traditional encyclopedias are typically curated by experts and editorial
teams who determine content based on their expertise, Wikipedia relies on
contributions from volunteers, enabling anyone to edit through open
collaboration, thus making it accessible to everyone. To maintain content
reliability and quality within its open platform, the Wikipedia community has
set up rules and guidelines that mobilize the collaborative efforts of
contributors in creating, deleting, or improving articles. In the following
sections, we delve into the gender gap in content and the community,
highlighting how the Wikipedia community applies standards for the subjects of
biographies. We then discuss attempts made to bridge this gender gap in content
and the challenges encountered. Additionally, we explore how the standards for
subjects are applied in the deletion process and develop hypotheses to test if
there is a gender disparity in the application of these standards in deletion
consideration.

\subsection{Gender Gap in Content and Community}

The lack of representation of women in the biographies on Wikipedia has garnered
extensive attention and analysis in various studies. For instance, Reagle and
Rhue compared English Wikipedia with the Encyclopedia Britannica, finding a
significant lack of biographies of women on Wikipedia, despite its otherwise
broad coverage~\cite{reagle2011gender}. The aforementioned work of Konieczny and
Klein~\cite{konieczny2018gender} on the Wikidata Human Gender Indicator further
confirmed this disparity, showing that only 18.51\% of biographies on English
Wikipedia are about women. Similar results hold across various popular language
editions of Wikipedia, including French, Arabic, Polish, Italian, Russian,
German, and Spanish, with the percentage of biographies of women ranging between
17\% and 23\%~\cite{Humaniki24}. Even though it is challenging to place these numbers within perspective, lacking a ground truth about the completeness of these classes within society~\cite{patel2024estimating}, within specific fields of knowledge,
Adams~\emph{et al.}~\cite{adams2019counts} reported the limited representation
of female sociologists on Wikipedia, while Schellekens~\emph{et
al.}~\cite{schneider2012deletion} observed similar trends in the fields of
Physics, Economics, and Philosophy, suggesting the presence of ongoing
disparities across different fields. 

This gender disparity in biographical coverage is not limited to contemporary or
modern periods but extends into the historical record. A systematic analysis by
Laouenan~\emph{et al.}~\cite{laouenan2022cross} found that the gender gap is
particularly pronounced for individuals born between the 17\textsuperscript{th} and 18\textsuperscript{th}
centuries, with women representing only 5\% to 10\% of available biographies
from that period. 

These findings collectively highlight the ongoing challenges faced by women for
representation on Wikipedia, and have spurred an ongoing debate and
self-reflection within the Wikipedia community. Due to its open nature, the
content we see on Wikipedia necessarily reflects the interests and preferences
of its community, but the existing gender gap in biographies may also indicate
the possibility of different editorial standards being applied to women. Indeed,
Ford and Wajcman~\cite{ford2017anyone} argued that while Wikipedia aims to
encompass all knowledge, it has not adequately represented the knowledge of
everyone due to the lack of diversity in its community. In 2019, Katherine
Maher, then CEO of the Wikimedia Foundation, reported that women make up only
15--20\% of total contributors~\cite{Kandek23closing}. This underscores the need
for a greater understanding of how the Wikipedia community applies standards for
the subjects of biographies. 

On Wikipedia, notability is regarded by editors as the golden standard to
determine if a particular subject deserves its own
entry~\cite{wikipedia2024notability}. In particular, the notability
guidelines of biographies specify that individuals are deemed notable if they
have received substantial coverage in multiple reliable, published, secondary
sources independent of the subject~\cite{wikipedia2024notabilitypeople}. The content we see on Wikipedia results from the notability tests conducted during collective editing and deliberation processes. 

Several studies have tried to observe how the standard of notability is applied
to the subjects of Wikipedia biographies, though there is not an accepted
operational definition of it nor it is clear whether it is being applied
consistently. For instance, as proxies for the notability of a given individual,
Wagner~\emph{et al.}~\cite{wagner2016women} used the number of distinct language
editions having a Wikipedia entry about that person, as well as the number of
results returned by the Google search engine when searching the person by name.
They found that, on average, women on Wikipedia possess higher notability than
men, suggesting the presence of a ``Glass Ceiling'' effect: an invisible yet
significant barrier that prevents women from attaining significant levels of
visibility, recognition, or representation within the platform. This effect
implies that Wikipedia editors may inadvertently establish a more stringent
notability threshold for women, affecting their representation and
acknowledgment on Wikipedia. This effect holds for historical figures too, as
women face a higher threshold for recognition due to historically inadequate
documentation of their merits and limited coverage across
languages~\cite{wanggender23}.

In another study focusing on academics, Lemieux~\emph{et
al.}~\cite{lemieux2023too} took an intersectional approach to determine whether
the notability guidelines of Wikipedia are applied consistently across genders
and different racial backgrounds, and in particular when the label ``too soon''
is applied in the deletion process. This term is invoked within an AfD
deliberation to signal that an individual does not yet possess adequate coverage
in independent, reputable sources, and thus it is `too soon' to grant them a
Wikipedia entry~\cite{wikipedia2024TooSoon}.  Using again the coverage of
individuals in search engines as a proxy for their notability, the authors found
that biographies of female scholars with high media coverage were more likely to
be deleted with a ``too soon'' rationale, regardless of their racial
backgrounds. This was not the case for male scholars, suggesting inconsistent
enforcement of the ``too soon'' policy. 

Taken together, these studies indicate that there may be systemic biases present
in the notability assessment process of Wikipedia, particularly affecting women
and individuals from marginalized groups. However, the measures used to quantify
notability in these studies may not fully capture the way that Wikipedia editors
assess notability. This limitation arises for two primary reasons: first, each
measure, such as the number of language editions or search engine results, has
its own constraints and may only capture certain aspects of notability. For
instance, a higher volume of search engine results does not guarantee reliable
sources that accurately reflect the notability of a person. Second, even if we
could precisely quantify notability based on Wikipedia guidelines
(i.e.,~coverage in multiple, independent, and reliable sources), this might
still not align with how editors actually assess notability in practice.
Schneider et al.~\cite{schneider2012deletion} found that notability is
frequently invoked in collaborative decisions concerning the deletion of
biographies during the Articles for Deletion (AfD) process. Since notability
assessments are an integral part of the editorial process, articles may face
nomination for deletion at any stage of development over their lifecycle.
Therefore, examining when the deletion deliberations occur and whether the
deletion outcomes vary based on the gender of the subjects in these biographies
can reveal if the notability standards are applied consistently.

\subsection{Feminist Interventions}

To counter these trends, a number of feminist interventions have been proposed
over the years. Wikimedia Foundation board member Rosie Stephenson-Goodknight
has emphasized that, along with increasing the number of women within the
community of editors, it is especially crucial to improve the coverage of
content about women~\cite{Kandek23closing}. In line with this idea, and as a
precursor to it, the ``Women in Red'' (WiR)~\cite{wikipedia2024WIR} WikiProject
was established in 2015, with the goal to highlight women who had previously not
been prominently featured or acknowledged on the platform, by creating new
biographical entries about them. The project owes its name from a feature of the
wiki software. In Wikipedia and other wikis, red links in the text of articles
indicate that they are linked to pages that either do not exist or previously
existed but have been deleted. Adding red links to text during the editing of
the articles essentially encourages other editors to create new pages and turn
the red links into blue. Since its inception, the Women in Red community sought
to turn red links pointing to the missing biographies of women into blue,
successfully increasing the percentage of biographies of women from 15\% to
19\%~\cite{wikipedia2024WIRmetric}.  Aside from Women in Red, several other
feminist movements like ``Art + Feminism'' and ``500 Women Scientists'' paved
the way to bridging the gender gap~\cite{langrock2022gender}. 

However, despite all these attempts, biographies of women still continue to face
deletion challenges. Tripodi~\cite{tripodi2023ms} studied the Women in Red
project and identified several instances in which notable women had been
nominated for deletion shortly after their creation. She also noted that content
about notable women, even when added and protected by community efforts like
Women in Red, can still face prompt deletion. Aside from coordinating and
participating in events aimed at enhancing the representation of women on
Wikipedia, editors also have to allocate significant time to ensure their articles
remain intact and survive deletion attempts~\cite{tripodi2023ms}, which adds
more burden and emotional labor to the fight for bridging the gender gap.

\subsection{Deletion Process}\label{sec:deletion_process}

Schneider~\emph{et al.}~\cite{schneider2012deletion} identified four different
deletion processes in Wikipedia: Speedy Deletion (CSD), Proposed Deletion
(PROD), Proposed Deletion for Biography of Living People (BLPPROD), and Articles
for Deletion (AfD). \emph{Speedy Deletion} covers all situations where deletion
is clearly the most appropriate, and least contentious, action, such as
vandalism, hate content, duplicates of existing pages, etc.; \emph{Proposed
Deletion} applies to situations that do not fall under the CSD guidelines but
for which no opposition to the deletion is expected, like articles about
promotional products; \emph{Proposed Deletion for Biography of Living People}
applies to biographies of living individuals that lack referenced merits; and
finally, \emph{Articles for Deletion} applies to all those remaining case where
editors are called to evaluate subjects where there is uncertainty or a lack of
consensus regarding their notability. 

What distinguishes AfD from other processes is its focus on notability
assessment, achieved through extensive discussion and deliberation. Editors can
nominate articles for potential deletion based on a number of notability
concerns. Once nominated, an entry is listed on a web board where self-selected
editors are invited to provide comment about any of the nominated entries.
Discussants typically suggest a range of actions, the most common of which can
result in either the deletion of the article (\emph{Delete}), the replacement of
the entire entry with a pointer to an existing entry (\emph{Redirect}), and the
merging of the entry with an existing entry (\emph{Merge}). If no consensus is
reached, or notability is affirmed, the page is left as is (\emph{Keep}). After
a period of about one week an administrator editor reviews the discussion, 
implements the decision, and closes the discussion.

Due to the self-selected nature of AfD discussion bodies, several prior studies
have focused on how the group composition of these deliberation affects their
outcomes~\cite{wang2023social, javanmardi2019s, mayfield2019analyzing,
taraborelli2010beyond, tasnim2021characterizing}. Taraborelli and
Ciampaglia~\cite{taraborelli2010beyond} were among the firsts to show that the
editors who take part to deletion discussions tend to fall under two different
camps, known as the \emph{inclusionists} and \emph{deletionists}, based on their
tendency to deliberate in favor of either Keep or Delete outcomes. Later, Tasnim
Huq and Ciampaglia~\cite{tasnim2021characterizing} used Gaussian mixture models
to categorize editors based on their past record of deliberation. They found
evidence for a more fine-grained taxonomy with a distinction between strong and
weak Inclusionist\,/\,Deletionists, and showed that the composition of the group
of discussants in terms of this group membership is a strong predictor of the
outcome of an AfD. This suggests that AfD deliberations may have a hard time to
reach a consensus when editors from these polarized groups are part of the
discussion, thus likely leading to longer, and more debated deliberations. 

\subsection{Gender Disparity in Deletion Process}

A number of studies have focused on the role of deletion deliberations in the
gender gap on Wikipedia. Worku~\emph{et al.}~\cite{worku2020exploring} examined
whether content preferences of editors for gendered content influence the
outcomes of deletion deliberations. The study found no evidence of systematic
biases in deletion decisions of articles that are of interest to men and women.
However, the authors did not consider factors influencing the selection of
articles for deliberation, like at what stage in the editorial cycle of an
article the nomination for deletion tends to occur. This is a crucial question,
since deletions are part of the regular collaboration process of editorial
development of the encyclopedia. Wikipedia editors choose a topic, start writing
articles, and gradually develop the articles by adding content and references.
Meanwhile, the other editors use the information they have access to (either
present in the the entry itself or in external sources) to assess the notability
of the entry subject, and based on this information argue for its inclusion or
deletion in case of an AfD. Articles nominated prematurely in this editorial
cycle may not have enough time to develop and thus to aggregate enough
notability information.

Relating notability discussions in AfD to the gender gap, in addition to the
aforementioned analysis of WiR, Tripodi~\cite{tripodi2023ms} also conducted a
comparative analysis of discussions regarding the deletion of biographies of
women and men. She analyzed the AfD discussions of 22,174 biographies nominated
for deletion between 2017 and 2020. She found that biographies of women were more
likely to be nominated for deletion than those of men, even though they were
also more likely to be kept (i.e.,~to be found to be notable) as a result of the
subsequent AfD deliberation. This is evidence that women are being
`miscategorized' as less notable than men. However, in this study, there is no
way to determine at which stage articles were nominated after their creation. This
means that the study did not account for the varying survival rates of articles
before their nomination. 

Consistent with these findings, Martini~\cite{martini2023notable} showed that
the women who are the subject of nominated biographies are more often called
into question and discussed more controversially in AfD, suggesting that the
notability of women is closely monitored and more vigorously challenged due to
biased views. These studies suggest that editors in AfD might promptly question
the notability of women in biographies, highlighting the importance of
understanding what factors may affect the chance of nomination for deletion over
the lifespan of the articles.

\section{Hypothesis Development}
\label{sec:hypotheses}

There are multiple reasons why it is important to estimate the chances of an
article being nominated for deletion over its lifespan. First, several studies
have shown that the biographies that have been on Wikipedia for a long time tend
to undergo multiple edits, revisions, and reviews by editors, leading to an
incremental improvement in quality with collaboration over time~\cite{zhang2020mining,flekova2014makes,wang2020assessing}.
Consequently, this suggests that biographies take time to develop fully, with a
richer set of references and content. Second, the efforts of the Women in Red
community and several feminist interventions have led to a significant increase
in biographies about women in recent years. Compared to entries that have
existed for many years, newly created articles tend to be shorter and less
developed, and thus face greater challenges and undergo more scrutiny for
quality, often being considered for deletion. Thus, there may be a possibility
that the notability of women is questioned when their newly created biographies
are still in the development process. Therefore, inspired by Tripodi~\cite{tripodi2023ms}, we aim to test the following hypothesis:

\noindent\rule{\columnwidth}{0.4pt}

\noindent\textbf{H1}:~\emph{The biographies of women have a higher likelihood of being quickly considered for deletion compared to those of men in AfD.}

\noindent\rule{\columnwidth}{0.4pt}

In particular, we aim to address this question while accounting for additional
factors that could influence considerations for deletion, which we detail in the next set of hypotheses. 

Wikipedia has undergone significant changes and developments since its
inception, including shifts in structure, norms, policies, topical trends, and
editorial community engagement~\cite{heaberlin2016evolution,
ortega2012wikipedia}. Over the years, Wikipedia has expanded its coverage from
central encyclopedic topics to encompass a wide range of general and specialized
subjects. As the reach and popularity of Wikipedia have grown over time,
the Wikipedia community has gradually established stringent rules for entry to
maintain the quality of content and to prevent scams and
vandalism~\cite{keegan2017evolution}. 

As a result of the growing awareness toward the gender gap in coverage, the
number of biographies about women has increased in recent years. On one side,
there are efforts to include a broader range of women in the encyclopedia but,
on the flip side, debates are emerging in AfD about whether allowing them
compromises the standards of Wikipedia~\cite{martini2023notable}. Thus, there is
a possibility that entries about women created in recent years may have been
increasingly targeted for deletion. Furthermore, the trend in editor retention
has shifted since 2006, with a noticeable drop in participation from newcomer
editors~\cite{halfaker2013rise}. This change has significantly impacted the
level of engagement in deletion discussions within AfD over time, such as a
reduced participation in deletion discussions by newcomer
editors~\cite{Geiger11alternative}. One study, for example, has found that a
vast majority of newcomer article creators do not participate in Articles for
Deletion discussions regarding articles they have
started~\cite{Geiger11newuser}. 

To examine whether there is a gender
disparity in how quickly biographies are flagged for deletion on Wikipedia, we
test the following hypotheses:

\noindent\rule{\columnwidth}{0.4pt}

\noindent\textbf{H2(a)}:~\emph{Biographies created in recent years are nominated for deletion in AfD faster than in earlier years.}

\noindent\textbf{H2(b)}:~\emph{Biographies of women created in recent years  are
nominated for deletion in AfD faster than in earlier years.}

\noindent\rule{\columnwidth}{0.4pt}

When analyzing the timing of nomination for deletion of a biography, it is also
crucial to take another dimension into account, which is the \emph{living
status} of its subject, i.e.~whether the subject of the biography being
discussed is alive or deceased. Establishing the notability of living people
(BLP) is especially challenging, since subjects may be experiencing brief bursts
of notoriety, and the quality and availability of external sources may be
lacking or in flux. Being featured on Wikipedia also symbolizes status, and
individuals often strive to be listed on Wikipedia and invest significant time
and effort toward that goal, even though the standards of Wikipedia discourage
self-promotion and try to maintain objectivity by preventing conflicts of
interest~\cite{wikipedia2024conflictOfinterest}. Finally, there are additional
limits imposed on what can be covered about living subjects, the presence of
unflattering information about themselves on Wikipedia could lead some subject
to consider taking legal action. Indeed, Joyce~\emph{et
al.}~\cite{joyce2011handling} have shown that the AfD process can be a mechanism
to handle the biographies of living people when they are deemed to be
controversial. 

When it comes to gender, it is important to account for the living status of the
subjects for two reasons: first, among biographies of living people, roughly
25\% of them are women~\cite{Gray23Gender}, and this figure rises to $~50\%$
when focusing on professions other than athletes; second, over the past century,
awareness toward gender equality has become generally more common in society,
leading to a reasonable expectation of a higher representation of women among
notable figures worthy of inclusion in the
encyclopedia~\cite{konieczny2018gender}. Therefore, there is the possibility
that the proportion of women featured in articles about living people is
significantly higher compared to the overall ratio of women across all
biographies. To confirm whether being alive influences the likelihood of being
nominated for deletion more than gender, we test the following hypotheses:

\noindent\rule{\columnwidth}{0.4pt}

\noindent\textbf{H3(a)}:~\emph{The biographies of living people are nominated for deletion sooner in AfD.}

\noindent\textbf{H3(b)}:~\emph{The biographies of living women are nominated for deletion sooner in AfD.}

\noindent\rule{\columnwidth}{0.4pt}

Prior studies about gender and AfD have not provided insights into the
persistence of the gender gap in biographies of historical figures. Several
studies found that Wikipedia is heavily biased toward contemporary knowledge,
which means the number of historical figures is lower than the number of
contemporary figures in Wikipedia~\cite{laouenan2022cross, jatowt2016digital}.
Also, Wikipedia contributors have less tendency to create content related to the
past since collective or social memory decreases over
time~\cite{au2011studying}. However, Jatowt~\emph{et al.}~\cite{jatowt2019time}
discovered that articles about historical figures tend to be more comprehensive
than those about contemporary individuals. Particularly, they observed a decline
in the length of articles for individuals born after the 17\textsuperscript{th}
century. This raises questions about the risk of nominations for deletion of
historical figures in AfD, especially considering the lack of female historical
figures on Wikipedia~\cite{laouenan2022cross, wang2023social}. This
motivates us to test the following hypotheses:

\noindent\rule{\columnwidth}{0.4pt}

\noindent\textbf{H4(a)}:~\emph{The biographies of historical figures are nominated for deletion sooner in AfD.}

\noindent\textbf{H4(b)}:~\emph{The biographies of historical women are nominated for deletion sooner in AfD.}

\noindent\rule{\columnwidth}{0.4pt}

It is crucial to understand what happens to biographies after being nominated
for deletion in AfD discussions. Prior studies by Tripodi~\cite{tripodi2023ms}
and Martini~\cite{martini2023notable} have focused primarily on the binary
outcomes of deletion and inclusion in AfD debates. However, other outcomes, such
as the merging with existing articles or redirecting to different entries, are
also common in AfD processes. These alternatives, while not resulting in
deletion, do not necessarily mean that the affected biographies are preserved as
independent entries. Merging often results in an entire biography being absorbed
into another entry, and redirection in a biography becoming a subtopic within a
broader topic. This is especially important when considering the visibility and
recognition of biographies of women on Wikipedia, as many redirected or merged
articles may still reflect reduced prominence and independence. Additionally,
Martini~\cite{martini2023notable}, which focused on the German Wikipedia, showed
that debates concerning biographies of women tend to be longer and more
extensive. This highlights the need to examine the duration and likelihood of
all outcomes --- not just deletion --- to better understand how these dynamics
unfold in the AfD in English Wikipedia. That motivates us to test the following
hypotheses:

\noindent\rule{\columnwidth}{0.4pt}

\noindent\textbf{H5}:~\emph{AfD deliberations to delete entries about women are faster than those of men.}

\noindent\rule{\columnwidth}{0.4pt}

\subsection{Our Contribution}

In this study, we make three contributions:

\begin{enumerate}[label=(\emph{\roman*})]

    \item We employ survival analysis to explore potential gender
    disparities in the timing of nomination for deletion, covering the entire
    history of the AfD process from January 15, 2001, to November 3, 2023. This
    allows us, among other things, estimate the likelihood of survival from
    nomination for deletion over the entire history of Wikipedia.

    \item We contribute empirically to a better understanding of nomination for
    deletion patterns by examining various factors, including the
    characteristics of biography subjects such as gender, living status, and
    historical status, as well as the chronological ages of the Wikipedia
    project itself. These analyses provide insights into the dynamics of
    deletion nominations and their associations with different factors.

    \item Beyond the decision to nominate an article, we examine the
    aforementioned factors and their impact on the outcomes of AfD discussions.
    To achieve this, we employ a multi-state survival model, which captures both
    the risk of nomination as well as that of different competing
    events (i.e., the different outcomes of an AfD) following the nomination
    stage. 

\end{enumerate}

\section{Methodology}
\label{sec:methods}

Our analysis required us to collect data about the following pieces of
information:
\begin{enumerate*}[label=(\emph{\roman*})]

    \item the list of all existing biographical entries in Wikipedia;

    \item the list of all biographical entries nominated for AfD;

    \item the date of nomination for deletion of the biographies in AfD;

    \item the creation dates of both nominated and non-nominated biographies.

    \item the list of biographies about living people and deaths by decade;

    \item the gender, date of birth, and date of death of the subjects of all
    the biographies mentioned above; and finally

    \item the logs of conversations, recommendations and outcomes of AfD
    discussions.

\end{enumerate*}
We describe data collection for each of these in the following section. We then
briefly describe the main statistical framework (survival analysis) we employ
for our study, and how it can be extended to account for the presence of
multiple states and multiple competing risks (i.e.~different deliberation
outcomes).\footnote{The data and code to replicate all analyses is provided at: XXX (blinded for peer review).}

\subsection{Data Collection}
\label{sec:methods_data}

\subsubsection{Biographies selection and entry life cycle metadata}
\label{sec:methods_data_lifecycle}

Unfortunately, Wikipedia lacks a single source of truth to reliably identify its
biographical entries. Instead, the biographical status of an encyclopedic entry
must be inferred either from available metadata or via manual coding. For
example, Tripodi~\cite{tripodi2023ms} relied on a mix of metadata (deletion
sorting lists maintained by AfD participants, plus entries about women tagged by
the Women in Red project) and manual coding to identify biographical content and
its associated gender. However, this approach is not scalable beyond a certain
size. Because our goal is to study the deletion process in its entirety, the
dataset we employ for our analysis, which comprises of a total of 1,975,779
biographies with complete information on their entry life cycle (i.e.,~time of
creation, nomination, and deletion), is a sample of the full set of Wikipedia
biographies, compiled starting from two main sources of metadata.

First, we used Quarry~\cite{wikipedia2024quarry} to query the Wikipedia database
for all entries marked, by means of a category link, as within the scope of
WikiProject Biography~\cite{wikipedia2024wikiprojBio}, the main community effort
devoted to maintaining biographical content on the English Wikipedia. This query
yielded 2,017,869 biographical entries (henceforth referred as the `WP-Bio'
set). Second, we queried Wikidata for all the records with the Q5 property,
which is used to identify instances of the `Human' class within the Wikidata
ontology~\cite{wikidata2024q5}. This yielded a separate set of 2,050,178 records
(the `Q5' set). Note that these two sets differ in their composition. The former
includes only \emph{extant} Wikipedia biographical entries i.e., either because
(\emph{i}) never nominated, or (\emph{ii}) nominated and kept or (\emph{iii})
nominated, deleted, and finally recreated at a later time; the latter
includes, besides extant biographies, also subjects for which either an entry in
Wikipedia has never been created, or for which, at a certain point, a Wikipedia
entry had been created but had been later deleted, either because nominated for
AfD or other deletion process (see Sec.~\ref{sec:deletion_process}).  

Another challenge in defining our sample of biographies has to do with the fact
that the relevant metadata needed to describe the life cycle of an entry
(creation, nomination, deletion, etc.) is not recorded in any single location.
Rather, it needs to be reconstructed from a snapshot of the Wikipedia database.
This process is necessarily incomplete and required us to make certain
simplifications.

If an entry has never been nominated for deletion, then the only relevant piece
of information regarding its life cycle is the timestamp of creation. This
information is stored in the `page' table of the Wikipedia database, and we
collected it using the Wikipedia REST API. If an entry has been nominated, the
way to determine information about its AfD depends on the outcome of the AfD
itself: for nominated entries that have been kept or redirected, we apply the
same method used for those never nominated (i.e.,~we queried the `page' table
through the Wikipedia API); instead, for those that have been deleted or merged,
we need to reconstruct this information from the `archive' table of the
Wikipedia database, which is the staging area where content is stored after
deletion in Wikipedia. 
  
For this latter step, to identify all biographies nominated (and later deleted)
in the AfD process, we again used Quarry to extract 513,712 AfD deliberations
that took place between January 15, 2001 and November 3, 2023. Note this is the
\emph{complete} set of all AfDs (i.e., not just those of biographies). To filter
out AfDs of non-biographical content, we parsed the entry title from the title
of the AfD discussion page, and matched it against the aforementioned set of
Wikidata `Human' records, which yielded a total of 87,873 biographies nominated
for AfD deliberation. We then joined this list of AfDs against the `archive'
table to find all biographies deleted due to AfDs (i.e.,~either as result of a
`Delete' or `Merge' deliberation), and extracted the timestamp of their
creation.

Next, to estimate when the AfD nomination occurred (regardless of its outcome),
we also extracted the timestamp of creation of the AfD discussion page itself,
again querying the `page' table in the Wikipedia database.

Finally, during this process, we came across a considerable number of entries
(28,244) that had been recreated after having previously been deleted in
an earlier AfD (i.e.,~whose creation timestamp from the `page' table was
\emph{after} their timestamp of nomination). To resolve this discrepancy, and
keep our multi-state modeling framework simpler, we queried the `archive' table
to find the original creation date and discarded the re-creation date. This
approach ensured that each nominated entry appeared only once in the dataset
(i.e.,~for their first nomination). This procedure reduced the number of AfDs to
84,366.

\subsubsection{Deletion Discussions Metadata}
\label{sec:methods_data_discussions}

We then parsed the contents of each AfD discussion to extract the title of the
discussed entry, the rationale provided for nominating the entry, the final
outcome of the deliberation, and the timestamp of the closing of the discussion.
In addition to these, we also collected variables describing the size of the
deliberation, such as the number of participants, the number of messages posted,
and the average number of words per message. Because the language used to the
describe the possible outcomes of an AfD varies greatly, we consolidated the
labels based on the classification of editorial outcomes described in
Table~\ref{tab:outcomes}.

\begin{table}[h]
\centering
\caption{Classification of editorial outcomes of AfD}
\label{tab:outcomes}
\small
\begin{tabular}{cp{4.5in}}
\toprule
\textbf{Outcome} & \textbf{Parsed Outcome}\\ 
\midrule
Delete & Delete, Deletion, Deleted, Speedy Delete, Snowball Delete, Strong Delete, Weak Delete, Soft Delete, Speedied, Remove, Erase, Not Include, Not Keep, Copyvio, A7, Delete G7\\
\midrule
Keep & Keep, Kept, Inclusion, Include, Speedy Keep, Snowball Keep, Strong Keep, Weak Keep, Soft Keep, Userfy, Userfied, No-consensus, No Delete, Moot, Move, Procedural closure, Close, Withdraw\\
\midrule
Redirect & Redirect to, Redirect into, Redirect with, Redirected\\ 
\midrule
Merge & Merge, Merged, Merge all, Merge to, Merge into, Merge with\\
\bottomrule
\end{tabular}
\end{table}

\subsubsection{Vital Records and Historical Classification}
\label{sec:methods_data_vital}

Similarly to the entry life cycle information, we were confronted with several
challenges when collecting vital records information about the subjects of the
biographies. Even though many biographical entries in Wikipedia include some
vital record information (like date of birth or and death), we decided to rely
exclusively on Wikidata, which provides this information in a more structured
format.

Therefore, we accessed Wikidata, both via its SPARQL endpoint and its REST API,
to collect gender, date of birth, and date of death of the members of the Q5
set. We then matched individuals in Q5 to their respective Wikipedia entry (if
any) using its Wikidata item identifier (`wikibase\_item') recorded in the
Wikipedia database. This join resulted in a subset of 1,965,822 extant
biographical entries (from the `WPBio' set) matched to their Wikidata records
(`Q5' set) with complete vital records information. In matching these two sets,
we found that only 52,047 biographies could not be matched to a Wikidata entry.
For deleted entries, since they are not included in the WPBio set, we instead
matched members of the Q5 based on the title extracted from the `archive' table.

After removing entries with other genders, in this final dataset
($N=1,975,779$), we found 1,584,817 (80.21\%) men and 390,962 (19.78\%) women
(see Table~\ref{tab:dataset} for a full breakdown). Among them, we have the
dates of birth for 90.6\% of the subjects, dates of death for 40.5\% of the
subjects, and both pieces of information for 39.4\% of the subjects. 

To determine whether an individual is a contemporary or historical figure, we
compare their date of birth against a conventional cutoff: if the subject was
born before 1907 (the year of birth of the known oldest living person at the
time of our study~\cite{wikipedia2024Oldest}), we categorize the biography a
`historical', based on the idea that nobody categorized as such should still be
alive. Else, if the date of birth is after 1907, we categorize the subject as
either `contemporary alive' or `contemporary dead' based on whether a date of
death is recorded in our data. For subjects who lacked a date of birth but had a
date of death, we estimated their date of birth by subtracting the average life
expectancy of the remaining subjects (70 years) from their date of death. Since
the choice of this cutoff is somewhat arbitrary, as a robustness test we also
tested alternative cutoff dates and repeated our main survival analysis with
those, finding results that are qualitatively similar to the one presented later
in the main text, see Appendix~\ref{sec:Appendix}.

Finally, a small fraction of the subjects (8\%) lacked both the date of birth
and death information in Wikidata. To find whether they belonged to the set of
historical or contemporary figures, we used PetScan~\cite{wikipedia2024petscan}
to match the titles of those entries against the lists provided by the Wikipedia
categories on `Living People' and `Deaths by decade'. This process left 25,894
biographies (1\% of the Q5 set) for which we could not find either a birth or
death information. We manually checked a random sample of 100 of them, and found
they were all alive at the time of study, hence we include all of them in the
final dataset as `Alive'.

To assess the accuracy of this heuristic classification, we manually reviewed a
random sample of 500 biographies, verifying whether the subjects were
historical, contemporary dead, or alive. We found that the accuracy of the
information extracted from Wikidata and Petscan was 98\% overall (100\% for
gender, 98\% for living subjects, 96\% for contemporary deceased, and 99\% for
historical figures).

\subsection{Survival Analysis}
\label{sec:methods_survival}

Survival analysis is a statistical framework designed to estimate the duration
until a particular event, such as a machine failure or the death of a patient
under treatment, and for identifying potential factors influencing this
likelihood. In this study, we apply this framework within the context of
Wikipedia biography deliberations for deletion to gain insight into various
aspect of the process, such as the time to nomination and the duration of the
ensuing deliberation. Fig.~\ref{fig:diagram} provides a schematic description
of the events that characterize this process, and of whose duration we are
interested in. After its creation, any entry remains the \emph{created} state
until it is \emph{nominated} for deletion. A deliberation then determines which
of four possible editorial outcomes (\emph{Delete}, \emph{Keep},
\emph{Redirect}, or \emph{Merge}) may happen. For each of the state transitions
in Fig.~\ref{fig:diagram}, we record the duration (in days) before the
transition happens. Because not all pages do get nominated by the end of our
observation window, it means that our observations about the duration of stay in
the `Created' state are affected by (right) censoring. That is, a biography is
considered censored if it was not nominated by the time of data collection
(November 3, 2023). Note that, in our dataset, all entries that have been
nominated also have a final deliberation outcome. In other words, none of the
deliberation duration observations are affected by censoring. 

In the following, we give a brief overview of the main elements of survival
analysis for multi-state models with competing risks. We begin by describing the
basic form of survival analysis for events of a single type (e.g. nomination),
and then generalize to the case of multiple event types, which is suited to
describe deliberations, with their multiple outcomes. All the analyses were
carried out using the `lifelines' Python package for survival
analysis~\cite{davidson2019lifelines} and the `survival' R package for the
multi-state analysis with competing risks~\cite{survival-book}.

\subsubsection{Survival Function $S(t)$}
\label{sec:methods_survival_st}

The probability of survival from, e.g., getting nominated for deletion after
time $t$. In our study, we use the Kaplan--Meier
estimator~\cite{kaplan1958nonparametric} to infer this probability, which is
given by the following equation:
\begin{equation}
  S(t) = \prod_{i \,: \, t \leq t_i}\left(1 - \frac{d_i}{n_i}\right)
\end{equation}
\noindent where $t$ is the time when at least one event (e.g. nomination)
happened, $d_i$ is the number of nominated entries at time $t_i$, $n_i$ is the
number of entries that have not been nominated yet at time $t_i$ or are
censored. 

\subsubsection{Hazard Function}
\label{sec:methods_survival_hazard}

We applied the Cox proportional hazards model~\cite{cox1972regression} to
estimate the risk occurrence of an event of interest (e.g. nomination) in terms
of a set of variables denoted by vector $\boldsymbol{X} = \left(X_1, X_2,
\ldots, X_n\right)$ that increase or reduce the risk $\lambda(t |
\boldsymbol{X})$ at time $t$ as given by the following model:  
\begin{equation}
  \label{eq:cox}
  \lambda\left(t \, | \,\boldsymbol{X}\right) = \lambda_0(t)\exp\left(X_1 \beta_1 + X_2 \beta_2 + \ldots + X_n \beta_n\right) = \lambda_0(t)\exp\left(\boldsymbol{X\beta}\right)
\end{equation}
\noindent where $\boldsymbol\beta = \left(\beta_1, \beta_2, \ldots,
\beta_n\right)$ is the vector of coefficients that measures the impact of the
variables in vector $\boldsymbol{X}$. The term $\lambda_0(t)$ is the baseline
hazard when $\boldsymbol{X} = \boldsymbol{0}$, i.e. all the variables are equal
to zero. Let $\boldsymbol{X}^{(i)}$ and $\boldsymbol{X}^{(j)}$ denote the
vectors describing the variables for the $i$-th and $j$-th biography, then the
hazard ratio $\mathrm{HR}$ between the two is 
\begin{equation}
\mathrm{HR} = \frac{\lambda\left(t \, | \, \boldsymbol{X}^{(i)}\right)}{\lambda\left(t \, | \, \boldsymbol{X}^{(j)}\right)} 
                  =\frac{\lambda_0(t)\exp\left(\boldsymbol{X}^{(i)}\boldsymbol{\beta}\right)} {\lambda_0(t)\exp\left(\boldsymbol{X}^{(j)}\boldsymbol{\beta}\right)}
                  = \exp\left(\left(\boldsymbol{X}^{(i)} - \boldsymbol{X}^{(j)}\right) \boldsymbol{\beta}\right)
\end{equation}
We fit the regression model of Eq.~\ref{eq:cox} to estimate the coefficients in
vector $\boldsymbol{\beta}$. We report these estimates as the associated
$\log\left(\mathrm{HR}\right)$ associated to each $\beta_i$, relative to
$\boldsymbol{X}^{(j)} = \boldsymbol{0}$. Negative values of $\beta_i$ correspond
to a reduction in the risk of occurrence for a unit change of variable $X_i$
(all else equal), while positive values to an increase in risk. 

\subsubsection{Multi-state and competing risks model}
\label{sec:methods_survival_multistate}

Since AfD deliberation can lead to multiple competing outcomes (Delete, Keep,
Redirect, and Merge), we use the Cumulative Incidence Function
($\mathrm{CIF}$)~\cite{gray1988class} to estimate the marginal probability for
each competing event, and thus to measure the impact of the above variables
(e.g. gender or status) on the overall deliberation, as given by the following
equation:
\begin{equation}
  \mathrm{CIF}_c(t_i) = \sum_{k=1}^i S(t_{k-1}) \cdot h_c(t_{k})
\end{equation}
\noindent where $c$ denotes the event type, $S(t_{k-1})$ is the Kaplan-Meier estimator of survival up to time $t_i$, and $h_c(t_k)$ is the hazard function for event type $c$ at any prior time $t_k$. 

\subsubsection{Regression variables}
\label{sec:methods_survival_variables}

In all regression analyses, we encode `gender' as a categorical value, with two
levels --  man ($0$) and woman ($1$) -- we always report the coefficient
associated to being a woman. Likewise, we use the vital records information
defined above (see Sec.~\ref{sec:methods_data_vital}) to define a categorical
value representing the vital status of a subject. In the following we report
coefficients for both the `historical' and `contemporary dead' levels. Finally,
to account for the historical evolution of Wikipedia (hypotheses H2(a) and
H2(b)) we defined a interval variable corresponding to the `age' of Wikipedia
(in years) at the time of creation of an entry. 

Finally, for the competing risks analysis, since we consider only entries
nominated for deletion, we also include in the regression additional variables
to describe the deliberation itself. These variables include the number of
participants, the number of messages posted, and the average message size (in
words).

\section{Results}
\label{sec:results}

Table~\ref{tab:dataset} shows a breakdown of the number of entries by gender,
nomination status, and living status of the subject. The date of birth of
subjects ranges from 7999~BCE to 2022~CE. In our dataset, the proportion of
women among nominated article (25\%) is greater than among all articles ever
published (20\%), which is consistent with prior work~\cite{tripodi2023ms}.

\begin{table}[h]
\centering
\caption{Data summary.}
\label{tab:dataset}
\small
\begin{tabular}{@{}cccccc@{}}
\toprule
\textbf{Gender} & \textbf{All} & \textbf{Nominated} & \textbf{History} & \textbf{Alive} & \textbf{Contemp. Dead}\\ 
\midrule
Women & 390,962 334 (19.8\%) & 21,473 (25.5\%) & 64,369 (11.9\%) & 276,975 (24.9\%) & 49,603 (15.3\%)\\
Men & 1,584,817 (80.2\%) & 62,893 (74.5\%) & 472,436 (88.1\%) & 836,696 (75.1\%)  & 275,637 (84.7\%)\\
Total & 1,975,779 & 84,366 & 536,805 & 1,113,671 & 325,240\\ 
\bottomrule
\end{tabular}
\end{table}

\subsection{Nomination for Deletion}

\begin{figure}
  \centering
  \includegraphics{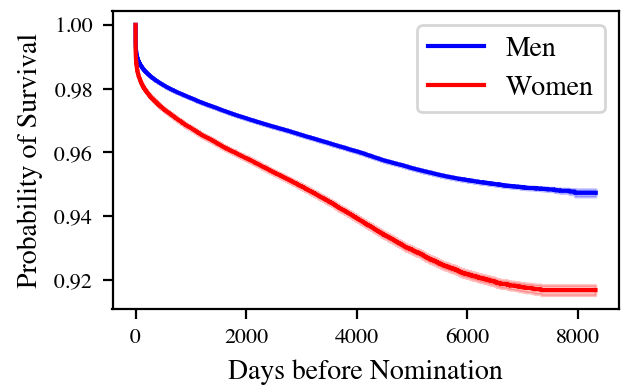}
  \caption{Probability of survival from nomination for deletion. The shaded area corresponds to the 95\% c.i.}
 \label{fig:kaplan_meier}
  \Description{The probability of survival of the biographies from nomination for deletion. The shaded area corresponds to the 95\% confidence intervals.}
\end{figure}

In Figure~\ref{fig:kaplan_meier}, the Kaplan-Meier curves for men and women in
biographies show early drops in survival, suggesting an ``infant mortality''
pattern. However, after this initial drop, the two lines exhibit different
slopes. Notably, the curve for women drops further down than that of men,
suggesting that biographies of women receive nominations for deletion more
rapidly than their male counterparts. However, these patterns could be explained
by various confounding factors described above (see Sec.~\ref{sec:hypotheses}),
and to account for these we employ the Cox
proportional hazards model to assess the risk of nomination controlling for a number of variables (see Sec.~\ref{sec:methods_survival_variables}).

\begin{figure}
  \centering
  \includegraphics[scale=0.85]{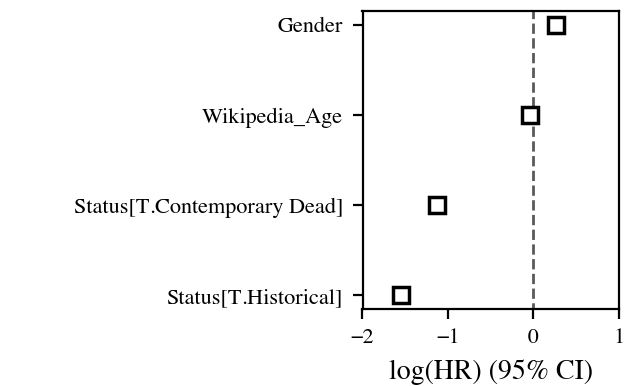}
  \includegraphics[scale=0.85]{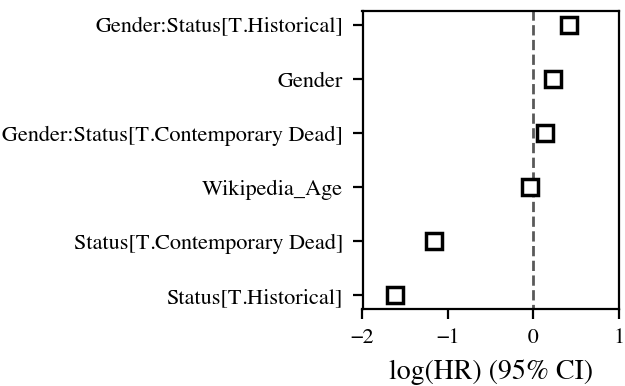}
  \caption{Cox regression analysis. Left: Baseline model; Right: the model with interaction terms between gender and status. Error bars represent robust standard errors and are all smaller than the data points.}
  \label{fig:cox_prop}
  \Description{Results of Cox proportional hazards models on the full dataset. First: Baseline model; Second: the model with interaction terms between gender and status. In both plots, error bars represent robust standard errors and are all smaller than the data points.}
\end{figure}

Figure~\ref{fig:cox_prop}~(Left) shows that only gender is positively associated
with nomination ($\beta=0.23$, $p<0.005$). When the subjects of biographies are
women (H1), the entries are nominated for deletion $26\%$ faster than those of men. Historical biographies are associated to the lowest risk ($\beta = -1.59$,
$p < 0.005$), indicating that historical figures (H4(a)) are considered for
deletion at an 80\% slower rate compared to contemporary living figures.
Finally, the age of Wikipedia has a negative but small association with
nomination ($\beta = -0.02$, $p < 0.005$), suggesting that entries created in
recent years are nominated at a slower pace than in the past (H2(a)).

Figure~\ref{fig:cox_prop}~(Right) shows the result of adding interactions terms
to the baseline Cox proportional hazards model. In particular with consider
interactions between gender and living status. A log-likelihood ratio test shows
that the model with interaction terms fits the data better than the baseline
model ($p<0.005$). We find that the interaction between gender and historical
status (H4(b)) is positively associated with nomination ($\beta=0.42, p<0.005$).
This suggests that although being a historical figure provides some protection
against nomination (H4(a) -- not supported), historical women still face a
disadvantage compared to historical men (H4(b) -- supported). In essence,
historical women experience a double penalty due to their gender and historical
status. 

\begin{figure}
  \centering
  \includegraphics{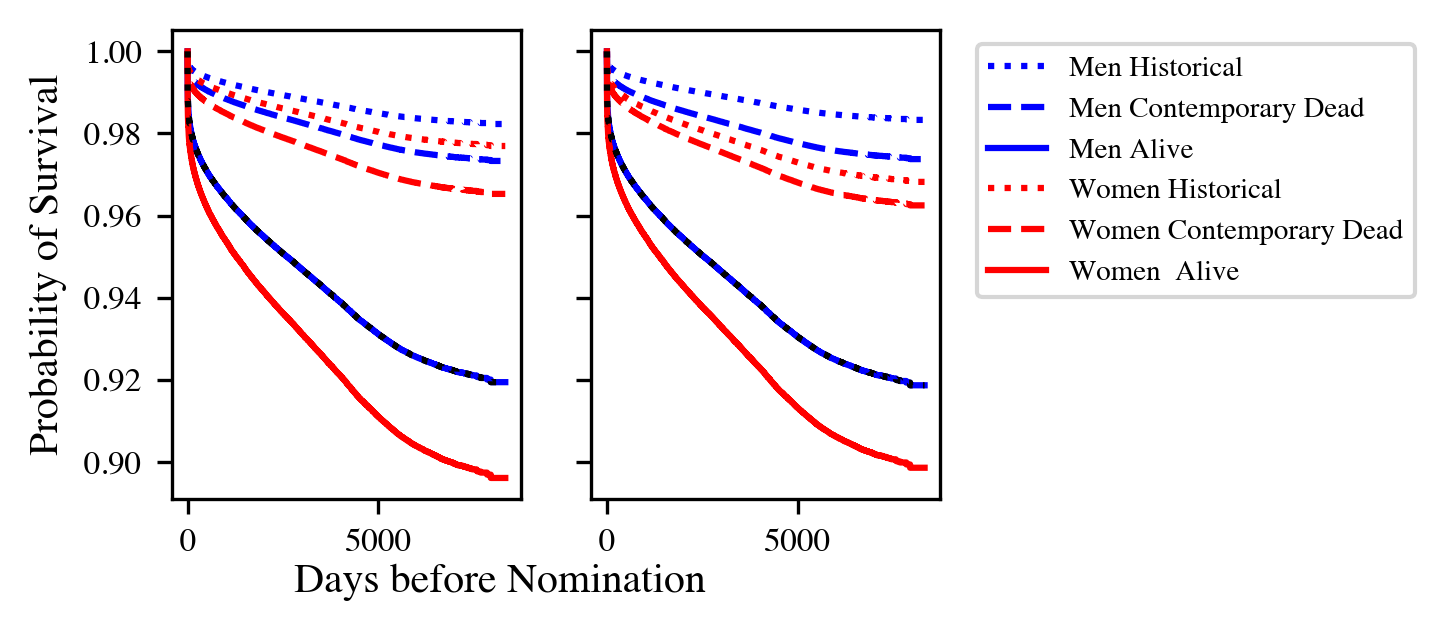}
  \caption{Marginal effects of gender and status. Left: Baseline model; Right: the model with interaction terms between gender and living status.}
 \label{fig:marginal_effect}
  \Description{Marginal effects of gender and status. Left: Baseline model; Right: the model with interaction terms between gender and living status.}
\end{figure}

To get a better sense of the effect of the interaction term, in the two panels
of Figure~\ref{fig:marginal_effect} we plot the partial effects of gender and
living status on the risk of nomination before and after incorporating the
interaction terms. In particular, Figure~\ref{fig:marginal_effect}~(Right) shows
that entries about dead women (either historical or contemporary) have both
higher risk of nomination than those of their dead men counterparts. Overall,
the entries about living people (H3(a)) tend to be nominated faster
compared to those of dead people (either historical or contemporary), with living women being the ones experiencing nominations the fastest (H3(b)).

\subsection{Deliberation Duration}

What are the chances of receiving a particular deliberation outcome, and are
deliberations about biographies of women different than those of men in terms of
their duration? To answer these questions, we start by focusing on the
biographies that reached the AfD deliberation stage. Table~\ref{tab:decisions}
provides a breakdown of these deliberations by the gender of the subject and
their outcome. 

\begin{table}[h]
\centering
\caption{Number of nominated biographies by gender and deliberation outcome.}
\label{tab:decisions}
\small
\begin{tabular}{@{}rccccc@{}}
\toprule
\textbf{Gender} & \textbf{Delete (D)} & \textbf{Keep (K)} & \textbf{Redirect (R)} & \textbf{Merge (M)} & \textbf{Total} \\ 
\midrule
Women & 9,432 (23.1\%) & 9,786 (28.1\%) & 1,078 (27.9\%) & 269 (27\%) & 20,565 (25.6\%)\\
Men & 31,388 (76.9\%) & 25,010 (71.9\%) & 2,792 (72.1\%) & 725 (73\%) & 59,915 (74.4\%)\\
Total & 40,820 & 34,796 & 3,870 & 994 & 80,480\\ 
\bottomrule
\end{tabular}
\end{table}

By far, the two most common outcomes for AfDs are deletion
($49.7\%$) and keeps ($42.4\%$) --- either due to consensus or lack thereof.
Across all outcomes, there are more deliberations about entries of men than of
women, owing to their higher prevalence in the encyclopedia. Furthermore,
conditional on nomination, men are deleted at slightly higher rates (52.4\%)
than women (45.9\%).


\begin{figure}
    \centering
    \includegraphics[width=\textwidth]{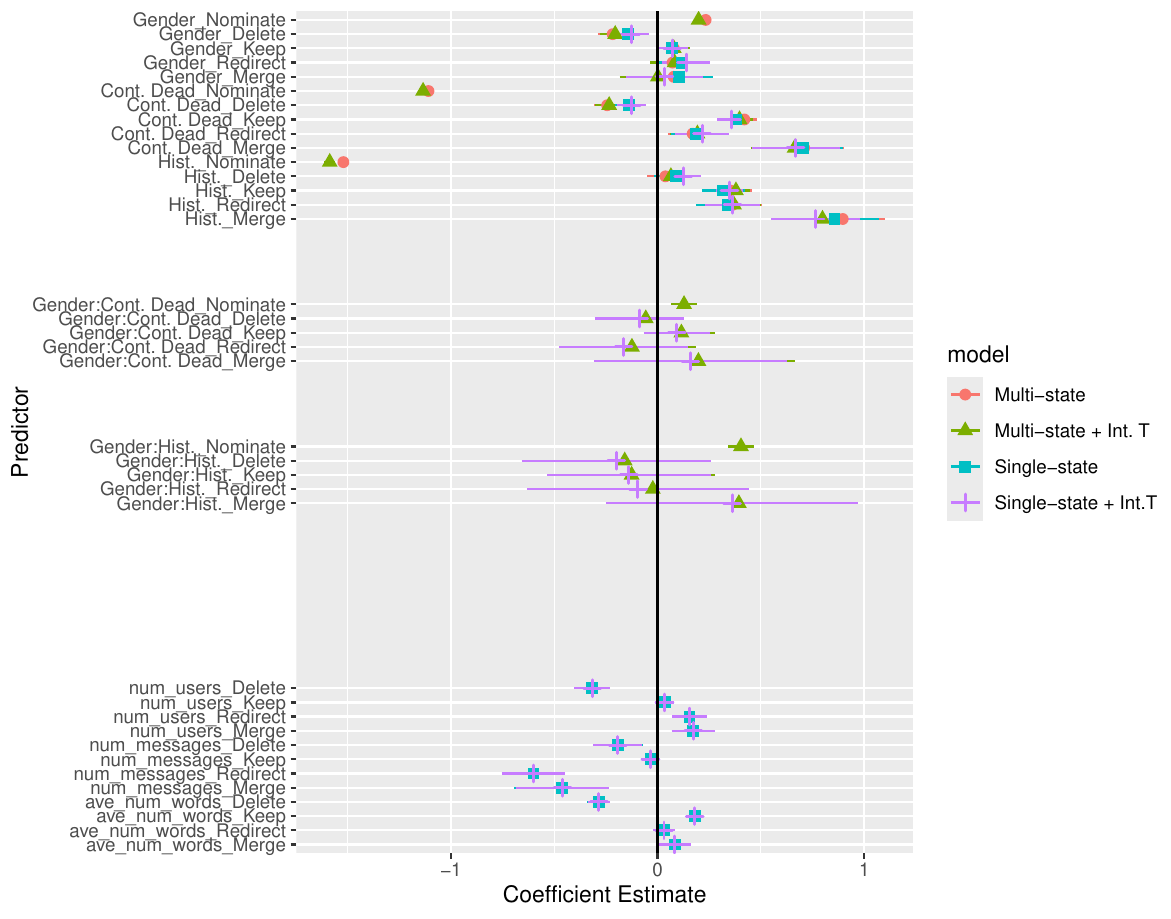}
    \caption{Survival analysis with competing risks. The `multi-state' models correspond to the full diagram in Fig.~\ref{fig:diagram}, while the `single-state' models correspond only to the final transition in the diagram (i.e. without the `Created' state). The error bars show the 95\% confidence intervals.}
    \Description{Survival analysis with competing risks. The `multi-state' models correspond to the full diagram in Fig.~\ref{fig:diagram}, while the `single-state' models correspond only to the final transition in the diagram (i.e. without the `Created' state). The error bars show the 95\% confidence intervals.}
    \label{fig:multistate}
\end{figure}

To better understand the duration of deliberation for different groups of
entries, we fit the full multi-state model with competing risks shown in
Figure~\ref{fig:diagram} to the full dataset of entries (i.e. nominated and
not nominated). Because this model includes both type of events (i.e.
nominations and deliberation outcomes), we can compare its results with those of
the vanilla survival model (i.e. without competing risks) used for the dataset
with nomination events only. 

Fig.~\ref{fig:multistate} summarizes the results, which for nomination events
align closely with those derived from the vanilla model of
Fig.~\ref{fig:cox_prop}. Consistent with it, the risk of being nominated for
deletion is lowest for contemporary dead (H3(a)) and historical (H4(a))
biographies, and is positive for biographies of women (H1). 

Moving to the effect of gender on the duration of deliberations for different
outcomes, we find that deliberations to delete entries about women are longer in
duration than those of men (H5), possibly suggesting a more disputed, and harder
to reach consensus in those cases. When it comes to biographies of deceased
individuals (both historical and contemporary), which have low risk of being
nominated, deliberations are typically brief across the board, suggesting that
once nominated, those cases are rather uncontroversial, with the exception for
deletions of contemporary dead individuals, which tend to be slow.

Adding interaction terms between gender and living status to this full model
confirms the results of Fig.~\ref{fig:cox_prop}~(Left) about the risk of
nomination --- historical and contemporary dead women pay an additional price in
terms of speed at which they are nominated. For the duration deliberation, we
mostly find non-significant results.

Finally, in Fig.~\ref{fig:multistate} we also fit a separate competing-risk
model only to the sub-dataset of nominated entries (Tab.~\ref{tab:decisions}).
This allows us to control for additional variables describing the size of the
deliberation (see Sec.~\ref{sec:methods_data_discussions}). This model still
entails competing risks, but is not a multi-state model anymore as it covers
only the final transition in Fig.~\ref{fig:diagram} -- hence it is a
single-state model with competing risks. We find that some indicators of larger
deliberations are associated to slower deliberations regardless of the outcome
(number of messages). More users and longer messages are associated to slower
deliberation for deletes, but slightly faster deliberations for other outcomes.
That said, controlling for size of deliberation does not result in major
differences compared to the full multi-state model (with and without
interactions between gender and living status).


\begin{figure}
 \centering
 \includegraphics{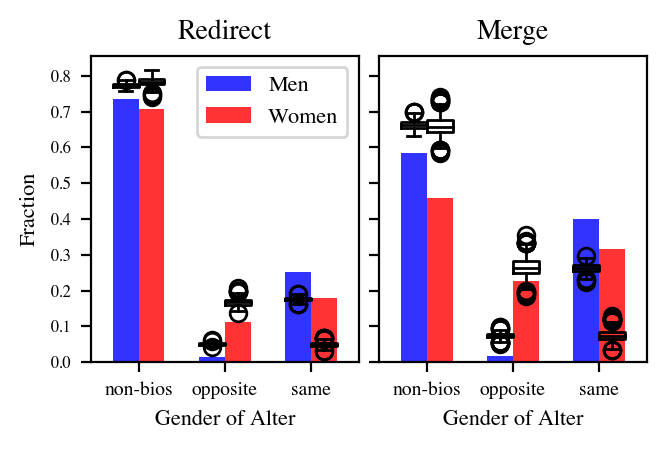}
 \caption{Fraction of merge / redirect outcomes by type of gender of alter for deliberations of both men and women. The black box and whisker plots refer to the same fraction under random reshuffling of the alter labels.}
 \label{fig:direction}
 \Description{Type of alters for merge / redirect outcomes.}
\end{figure}

The competing risks analysis shows that decisions to merge and redirect are
deliberated quickly, possibly suggesting these decisions are less contentious
and easier to reach, especially for deceased individuals. In both of these
outcomes, the nominated subject loses its status as an independent entry in the
encyclopedia, and is instead tied to the entry of an alter --- either a
biography or a non-biographical entry. This raises the question of what type of
alters these nominated entries are merged or redirect into.

Figure~\ref{fig:direction} shows the fraction of merges and redirects by gender
of biography and type of alter. For both biographies of men and women,
incorporation into a non-biographical alter is the most common forms of merge or
redirection, followed by same--gender alter. We find, however, a strong
asymmetry for merges and redirections into opposite--gender alters: the fraction
of men merging or redirecting into women alter is very low ($<2\%$), while the
reciprocal (i.e. women being merged into men) is nearly an order of magnitude
larger ($~20\%$). These observed patterns cannot be explained simply by the
presence of more biographies of men, as the frequencies obtained by random
reshuffling ($N=10^3$) of the three possible alter labels (man, woman,
non-biographical) significantly differ from the observed frequencies. 

\begin{figure*}
  \centering
  \includegraphics{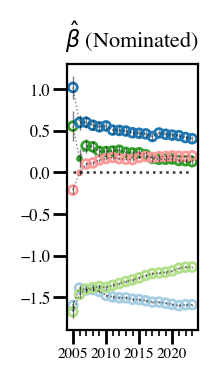}
  \includegraphics{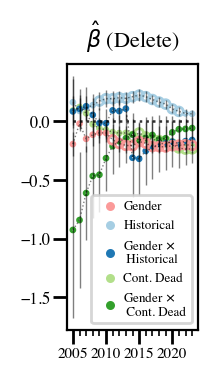}
  \includegraphics{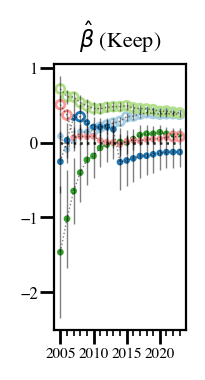}
  \includegraphics{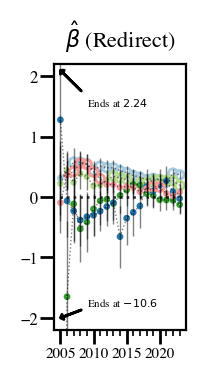}
  \includegraphics{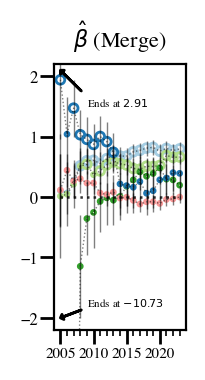}
  \caption{Retrospective analysis of multi-state model with competing events. Each figure shows the trend of the coefficient of each outcome over time. In all the figures, each data point corresponds to the coefficients of the Cumulative Incidence Function, fitted only on the data of articles created up to that year. The error bars represent robust standard errors. The black dash-dotted line corresponds to a coefficient value of zero.}
 \label{fig:retrospective_2}
  \Description{Retrospective analysis of multi-state model with competing events. Each figure shows the trend of the coefficient of each outcome over time. In all the figures, each data point corresponds to the coefficients of the Cumulative Incidence Function, fitted only on the data of articles created up to that year. The error bars represent robust standard errors. The black dash-dotted line corresponds to a coefficient value of zero.}
\end{figure*}

Finally, in a retrospective analysis of the multi-state survival model shown in
Figure~\ref{fig:retrospective_2}, we observe how the coefficients of variables such as gender and
living status change, as we consider observations windows of different sizes. This allows to assess how the risks for nomination and for the duration of deliberations evolved over the history of Wikipedia. Early on, the influence of
gender on nomination risk was negative (Figure~\ref{fig:retrospective_2}(a)),
but it steadily increased until 2006, and remained consistently positive
thereafter. Also, both historical and contemporary deceased women are at a
disadvantage from the very beginning and are still at risk.

Since 2010, deliberations to delete women have taken consistently longer than those of men. On the other hand, deliberations to delete historical figures tended to be fast, peaking around 2015, and have slowed down ever since. 

Deliberations to keep women have only very recently become faster than those of men, while those for keeping contemporary dead and historically have been consistently faster, suggesting they are less contentious than those of living people.

Decisions to redirect women used to be taken quickly until around 2010, but have since slowed down, unlike those for historical and contemporary dead people, which have consistently small but positive coefficients for both redirects and merges. For merges, deliberations of historical women used to be fast until the early 2010s but have since slowed down too.

Finally, despite being not statistically significant, we also find a slow down in duration for deliberations regarding historical women happening aorund 2014. Since the `Women in Red' initiative was established around that time (2015), it is possible that this intervention may have influenced the overall trends in AfD outcomes around that time.

\section{Threats to Validity}
\label{sec:threats}
Even though our analysis is based on the largest dataset of biographies
nominated in the AfD process collected so far, our coverage is still limited,
since the vast majority (68\%) of the entries nominated for deletion in the AfD
process are not present in our data. Thus, there might be a significant number
of biographies still missing in our analysis. Furthermore, as we focus only on
the AfD process, our analysis may miss entries deleted through other processes
as PROD or CSD. Furthermore, our gender labels come from Wikidata, but since
Wikipedia and Wikidata operate independently, its coverage of Wikipedia
biographies is far from complete. Our study thus inherits any potential bias
that may be present in the way gender information is annotated in Wikidata. 

\section{Discussion}
\label{sec:discussion}

Our study aims to provide an empirical foundation for the ongoing discussions
surrounding the gender gap in Wikipedia, particularly investigating whether
women in biographies are prone to nomination for deletion in AfD sooner than men
and how gender influences the deliberations of those AfD discussions. Table
\ref{tab:discussion} summarizes the results of our hypotheses.

\begin{table}[h]
\centering
\caption{Findings (\checkmark = supported; $\times$ = not supported).}
\label{tab:discussion}
\small
\begin{tabular}{@{}p{4.5in}c@{}}
\toprule
\textbf{Hypotheses} & \textbf{Conclusion} \\ 
\midrule
H1: The biographies of women have a higher likelihood of being quickly
considered for deletion compared to those of men in AfD. & \checkmark \\ 
H2(a): Biographies created in recent years are nominated for deletion in AfD
faster than in earlier years. & $\times$ \\ 
H2(b): Biographies of women created in recent years are nominated for deletion
in AfD faster than in earlier years. & $\times$ \\ 
H3(a): The biographies of living people are nominated for deletion sooner in
AfD. & \checkmark \\ 
H3(b): The biographies of living women are nominated for deletion sooner in AfD.
& \checkmark \\ 
H4(a): The biographies of historical figures are nominated for deletion sooner
in AfD. & $\times$ \\ 
H4(b): The biographies of historical women are nominated for deletion sooner in
AfD. & \checkmark \\ 
H5: AfD deliberations to delete entries about women are faster than those of men. & $\times$ \\ 
\bottomrule
\end{tabular}%
\end{table}

Our research reveals a tendency to prematurely scrutinize the notability of
women in the AfD at a higher rate than men. Since articles are continually being
developed, the earlier they are nominated for deletion, the shorter the window
for further development becomes. Thus, entries about women have lower chances to
develop and improve on Wikipedia. Our findings are also consistent with the
`miscategorization' hypothesis~\cite{tripodi2023ms}. Specifically, while
articles about women are more likely to be nominated for deletion at an earlier
stage compared to those about men, women are deleted at lower rate than men, and
those deliberation that decide on delete, take longer, presumably because more
disputed by editors. 

This suggests that editors may prematurely nominate articles about women based
on perceived lower notability, rather than thoroughly assessing the overall
importance of the subject, even when they meet the notability criteria of
Wikipedia with substantial media coverage~\cite{martini2023notable,
lemieux2023too}. However, the decisions to keep, redirect, or merge these
biographies point to a corrective mechanism within the AfD process, where
deliberation shifts towards properly evaluating notability. Moreover, while it
takes only one editor to nominate an article for deletion, saving it typically
requires collective deliberation. This highlights the vulnerability of the
biographies of notable women to premature deletion due to insufficient
assessment prior to nomination. It also indicates that the editors who nominate
these articles and those who argue for keeping them might interpret the deletion
policy differently.

Furthermore, our retrospective analysis shows that gender has long been a
significant factor in deletion nomination risk, even after the deployment of
feminist interventions like the ``Women in Red'' project. Moreover, we find that
historical women pay a double price when it comes to nominations as they are
disadvantaged for being both women and historical. Although contemporary living
individuals have a higher chance of being nominated for deletion, living women
face a greater risk of nomination than living men. Taken together, these
findings suggest that the notability of women is quickly undermined, giving them
fewer opportunities to enhance their Wikipedia presence. This phenomenon could
be attributed to the societal perception that men hold greater value and worth
than women \cite{berger1980status, eagly1982inferred}. 

Another explanation of why women are quickly nominated for deletion stems from
how they are described in Wikipedia. The entries of women are more likely to
contain language indicating their gender~\cite{tripodi2023ms}. The excessive use
of gendered terms such as ``female singer'', ``wife of'' or ``daughter of''
reinforces traditional gender roles and norms. This not only influences existing
notions of masculinity and femininity but also molds the identity of the
individuals being referenced~\cite{butler1990feminism}. This practice of
labeling content about women with gendered language establishes a heteronormative
structure, where a notable individual is automatically assumed to be male unless
specified otherwise. 

Using gendered terms in some specific cases, such as ``female president'' or
``female astronaut'' highlights the unique and pioneering aspect of the
accomplishments of women, which indeed aligns with the criteria for notability
of Wikipedia. However, contributors in AfD often argue that gender should not
determine notability, leading to a quick dismissal of women~\cite{martini2023notable}.  

Moreover, while this study lends support to the original `miscategorization'
theory, we provide a more complete picture of the full range of editorial
practices associated with it. In particular, we find an asymmetry in the choice
of alters into which merges and redirects occur, with opposite--gender
alters being much less common for men compared to women. This could be
because women are often portrayed as companions, partners, offspring, or
relatives of notable men, rather than being recognized as notable individuals in
their own right. As a result, women who are redirected or merged into articles
about men or non-biographical topics may continue to face reduced visibility as
independent subjects on Wikipedia. 

Conversely, the reverse pattern -- merging or redirecting men into entries about
women -- is rare, indicating that men are less likely to be positioned as
secondary figures or subtopics in biographies of women. This also suggests that
men are not typically viewed as mere helpers or subordinates in relation to
women.

Traditional historical narratives have long obscured, neglected, and distorted
the rich and diverse experiences of women throughout history
\cite{lerner1988priorities}. These narratives often focused primarily on the
achievements, perspectives, and roles of men, leaving women and their
experiences in the shadows. As a result, the voices and stories of women were
marginalized or omitted in both offline and digital media, including Wikipedia,
leading to an incomplete and biased understanding of historical events and
societal developments \cite{wherearethewomen, hamilton2017twenty,
edwards2015wiki}. The underrepresentation of women in historical and
contemporary records and sources contributes to a perception of lower
notability. This may prompt AfD contributors to undervalue women and take
premature actions based on that perception.

The gender disparity observed in this and other studies is challenging to
rectify and reflects a larger structural issue. However, Women in Red and other
feminist interventions exemplify collaborative efforts to address the gender gap
problem. From the retrospective analysis of the multi-state model, there is some
limited evidence of reversals in some of the trends around 2015, suggesting that
this and other feminist interventions may play a role in preserving biographies
of women. 

Our findings suggest that providing suitable time between creating and
considering deletion allows biographies of notable people to be well developed,
which is particularly beneficial for individuals for whom secondary sources
might be challenging to find. One possible way to do this could be to design a
recommendation interface for the AfD nominators that can help them assess the
notability of the subject before flagging for deletion. In the long term,
revising the Notability guidelines could also be a key step in addressing the
gender gap on Wikipedia~\cite{menking2021wp}. This revision should aim to
promote inclusivity and facilitate the representation of a wider range of
topics, including those related to women. 

Future work should explore how editors seek information about the subjects of
biographies on Wikipedia to assess notability, particularly focusing on gender
differences and types of reliable sources used as evidence. These efforts could
enhance our understanding of notability assessment processes and improve the
reliability and inclusivity of content on Wikipedia.

In conclusion, our study contributes valuable insights into the ongoing
discussions surrounding gender disparities and historical representation on
Wikipedia, highlighting the challenges and opportunities for fostering a more
inclusive and representative online encyclopedia.

\begin{acks}
This research is supported in part by the Wikimedia Foundation’s Research Fund
under Grant No.~G-RS-2303-12080. The authors would like to thank Victoria Van
Hyning, Sheena Erete, Wei Ai, Do Won Kim, Jay Patel, Marilyn Harbert, and Caro
Williams-Pierce for their generous feedback on an earlier version of this
manuscript.
\end{acks}

\bibliographystyle{ACM-Reference-Format}
\bibliography{references}
\newpage
\appendix
\section{Cutoff Date for Historical Figure}
\label{sec:Appendix}

As a robustness check, we considered two more alternative cutoff years in the
analysis: a) The start of the late modern period (1800), and b) The start of the
contemporary period (1945). We found that using these
alternative cutoff choices does not alter the conclusion of our hazard model, see Fig.~\ref{fig:cox_prop_1800}:

\begin{figure}[ht]
  \centering
  \includegraphics{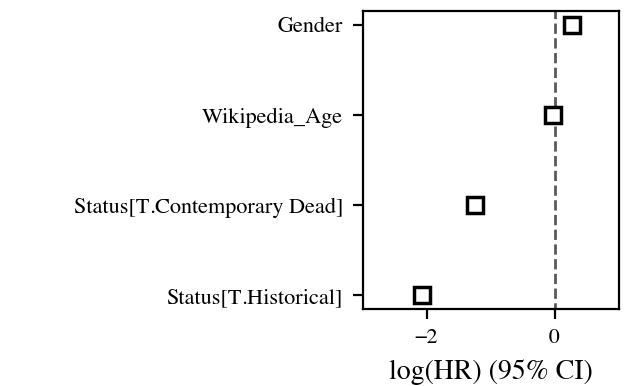}
  \includegraphics{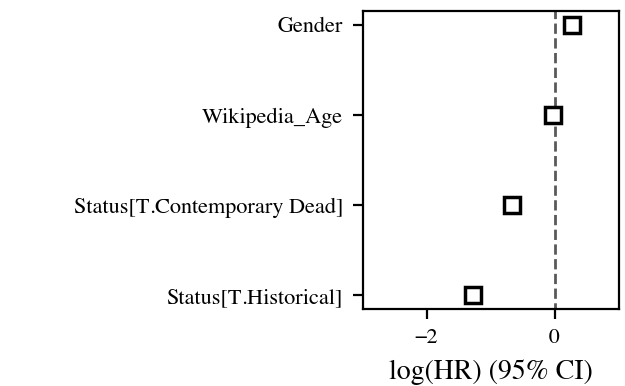}
  \caption{The results of the Cox proportional Hazards model across cutoff years. First: when cutoff year is 1800; Second: when cutoff year is 1945. In both plots, error bars represent robust standard errors and are all smaller than the data points.}
  \label{fig:cox_prop_1800}
  \Description{Results of Cox proportional hazards models on the full dataset. First: Baseline model; Second: the model with interaction terms between gender and status. In both plots, error bars represent robust standard errors and are all
smaller than the data points.}
\end{figure}
\end{document}